\documentclass[12pt]{article}

\usepackage{amssymb}
\usepackage{amsmath,epsfig,subfigure}

\newcommand{\la}{\label}
\newcommand{\be}{\begin{equation}}
\newcommand{\en}{\end{equation}}

\renewcommand{\vec}[1]{\boldsymbol{#1}}
\newcommand{\ii}{\textrm{i}}
\newcommand{\ee}{\textrm{e}}

\def\thirdt{{\textstyle\frac{1}{3}}}

\begin{document}

\title{Bending instabilities of soft tissues}

\author{Michel Destrade$^a$, Aisling N\'i Annaidh$^{a,b,c}$, Ciprian Coman$^d$ \\[12pt]
$^a$School of Electrical, Electronic, and Mechanical Engineering, \\
University College Dublin, Belfield, Dublin 4, Ireland\\[12pt]
$^b$UPMC Univ Paris 06, UMR 7190, \\
Institut Jean Le Rond d'Alembert, F-75005 Paris, France\\[12pt]
$^c$CNRS, UMR 7190, \\
Institut Jean Le Rond dÕAlembert, F-75005 Paris, France\\[12pt]
$^d$Department of Mathematics, University of Glasgow,\\ Glasgow G12
8QW, United Kingdom}

\date{}

\maketitle



\begin{abstract}

Rubber components and soft tissues are often subjected to large bending deformations ``in service''.
The circumferential line elements on the inner face of a bent block can contract up to a certain critical stretch ratio $\lambda_\text{cr}$ (say) before bifurcation occurs and axial creases appear. 
For several models used to describe rubber, it is found that $\lambda_\text{cr} = 0.56$, allowing for a 44\% contraction. 
For models used to describe arteries it is found, somewhat surprisingly, that the strain-stiffening effect promotes instability. 
For example, the models used for the artery of a 70 year old human predict that $\lambda_\text{cr} = 0.73$, allowing only for a 27\% contraction.
Tensile experiments conducted on pig skin indicate that bending instabilities should occur even earlier there. 

\end{abstract}



\newpage


\section{Introduction}


When an elastic block is subjected to severe bending, it is expected that its inner curved face will eventually become unstable.
This phenomenon is well captured by the theory of incremental nonlinear elasticity, which can account for the appearance of small-amplitude wrinkles, aligned with the axial direction of the large bending. 
For a block made of rubber, Gent and Cho (1999) argue that this buckling should occur once circumferential line elements on the bent face are contracted up to the critical stretch ($\lambda_\text{cr}$, say) of surface instability for a semi-infinite body.
In that latter case, the neo-Hookean model predicts (Biot, 1963) that $\lambda_\text{cr} = 0.54$, allowing line elements to be contracted by 46\%.
Analyses looking into the bending instability (instead of the surface instability) of neo-Hookean blocks show that the finite size of a block (as opposed to the infinite size of a half-space) introduces dispersion (and thus links the size of a block to the number of wrinkles) but has little effect on the amount of possible contraction, bringing $\lambda_\text{cr}$ from 0.54 to 0.56, irrespective of the dimensions of the block.

Now the neo-Hookean model is often used to describe solids undergoing finite but moderate deformations. 
In fact, as recalled in Section \ref{Basic_equations} below, it encompasses the most general incompressible solid of third-order elasticity, as well as the Mooney-Rivlin model for rubber,  when the deformation is a plane strain such as pure bending. 
For large strains however, the polymers chains in elastomers and the collagen fiber bundles in soft tissues align themselves with the direction of greatest stretch and their \emph{limiting extensibility} is strongly felt. 
This phenomenon is the so-called \emph{strain-stiffening effect}.
In biological soft tissues the stiffening occurs at markedly lower strain levels than in rubber-like solids: broadly speaking, rubbers can be stretched \emph{at least} 100\%, whilst soft tissues can be stretched \emph{at most} 100\%, before they stiffen.

There are two popular models of constitutive law to describe the strain-stiffening effect: 
the \emph{Gent model}, which introduces a limiting chain parameter ---and thus an upper bound for the range of possible stretches---, and the \emph{Fung model}, which exhibits an exponential increase of stress with respect to strain ---but no maximal stretch.
The two models are presented in Section 2, along with typical values for the stiffening parameters of soft tissues. 
In particular, we conducted tensile experiments on pig thoracic aortas and found good agreement with data published previously in the literature.
 
Then in Sections 3 and 4,  we show that when these two models are adjusted to account for the actual physiological values of mammalian arteries, they predict that bending instabilities appear early, at moderate amounts of bending. 
For example , the Gent model, with a stiffening parameter adjusted to describe the thoracic artery of a 70 year old human, predicts that $\lambda_\text{cr} = 0.73$, which means that circumferential line elements can be contracted by 27\% only, before instabilities arise.
Further, the two models predict that stiffer tissues such as skin should become unstable almost as soon as they are bent. 
To estimate the stiffening parameters in those cases, we conducted tensile experiments on pig skin, and present the corresponding data and curve fitting results at the end of section (\S 4).

These results suggest that other characteristics must be accounted for in the constitutive modeling of arteries and skin. 
In particular, other investigations e.g. (Destrade et al., 2008) show that the anisotropy displayed by biological soft tissues (as opposed to the isotropy of elastomers) plays a most significant role in stability studies.


\section{Basic equations}
\label{Basic_equations}



\subsection{Constitutive laws}


We are interested in hyperelastic strain energy densities 
($W = W(\lambda_1, \lambda_2, \lambda_3)$ where the $\lambda$'s are the principal stretches)
commonly used to model the behavior of strain-stiffening  incompressible solids.
One of the simplest and best prototypes available to do this is the \emph{Gent model} (Gent, 1996),
\be \la{gent}
W = - \dfrac{\mu J_m}{2} \ln \left( 1 - \dfrac{\lambda_1^2 +
\lambda_2^2 + \lambda_3^2 - 3}{J_m} \right),
\en 
where $\mu > 0$ is the shear modulus for  infinitesimal strains and $J_m>0$ is a stiffening parameter, associated with limiting chain extensibility (Horgan and Saccomandi, 2006). 
Hence a solid behaving according to this constitutive model cannot be stretched in uniaxial tension 
beyond a maximal stretch $\lambda_m$, say, which is the positive root of 
\be
\lambda_m^2 + 2\lambda_m^{-1} - J_m - 3 =0.
\en
Typically, rubbers attain maximal strains ranging from 5 to 15;
in contrast, biological soft tissues experience strain-stiffening effects much earlier.
For instance experimental data on young, healthy human (Horgan and Saccomandi, 2003) and bovine (Humphrey, 2003) arteries show that they can be 
stretched up to a maximum of $\lambda_m = 1.4$, or even $2.0$, while older, stiffer arteries can  be stretched only up to 20\% max ($\lambda_m = 1.2$).
Accordingly, in what follows we take $20 \le J_m \le 200$ \emph{as the typical range for rubbers},
and we take $0.4 \le J_m \le 2.3$ \emph{as the typical range for arteries}. 
Other biological soft tissues belong to that latter range, including caterpillar muscle (Dorfmann et al., 2007), rabbit muscle (Davis et al., 2003), dura mater (Maikos et al., 2008), brain tissue (Franceschini et al., 2006), etc.

Another popular model in the biomechanics literature, which reflects a less pronounced 
strain-stiffening effect than the Gent model, is the \emph{Fung model},
\be \label{fung}
W = \dfrac{\mu}{2 b} \left[\ee^{b(\lambda_1^2 + \lambda_2^2 + \lambda_3^2 - 3)} - 1\right],
\en 
where $\mu > 0$ is the shear modulus at small strains and $b>0$ is the stiffening parameter.
For a human young thoracic artery, $b \simeq 1.0$, and for an older, stiffer artery, $b \simeq 5.5$, typically (Horgan and Saccomandi, 2003).
Accordingly, we take $1.0 \le b \le 5.5$ \emph{as the typical range for arteries} when using the Fung model.
This range is also inclusive of Fung model fitting for human cerebral veins 
($b \simeq 1.7$) and cerebral arteries ($b \simeq 4.4$), see  Ho and Kleiven (2007).

\begin{figure}
\centering \mbox{\subfigure{\epsfig{figure=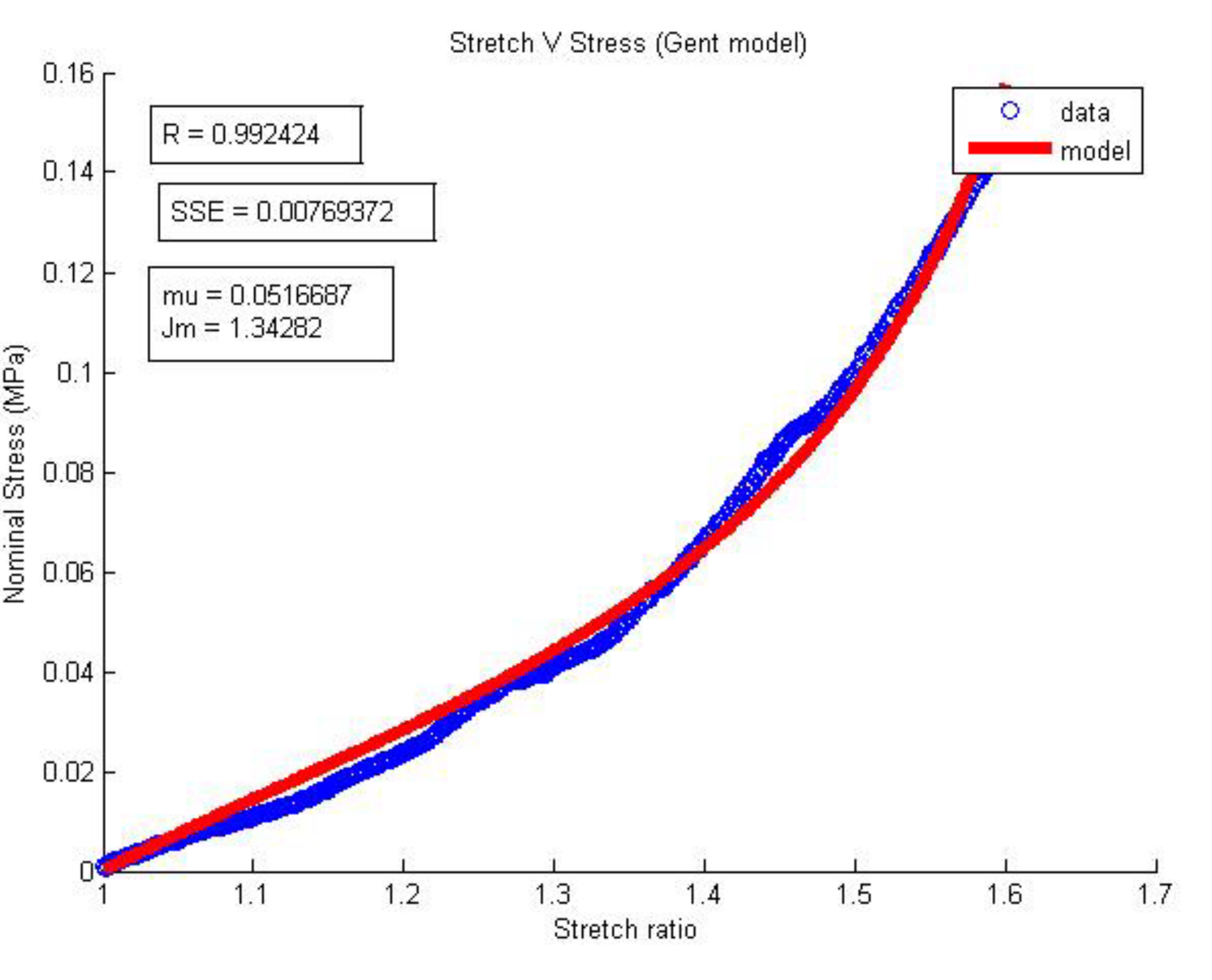, width=.45\textwidth}}}
  \quad \quad
     \subfigure{\epsfig{figure=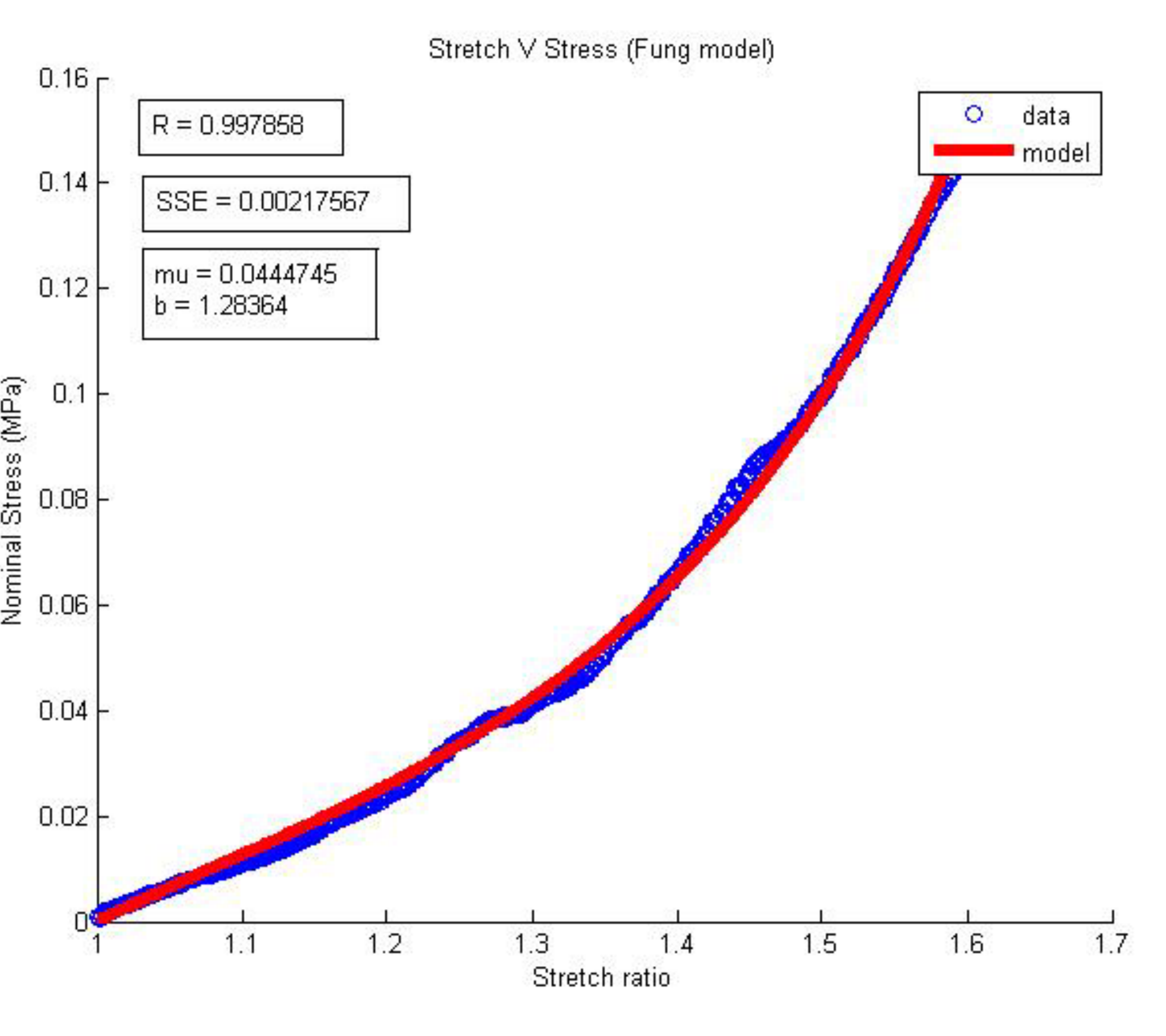, width=.45\textwidth}}
 \caption{Tensile test on a pig thoracic aorta: nominal stress versus stretch ratio. The thick line corresponds to experimental data and the thin line, to the curve fitting using the Gent model (on the left) and the Fung model (on the right). 
 Curve fitting analysis gives $J_m \simeq 1.4$ with the Gent model  and $b \simeq 1.3$ with the Fung model.}
 \label{fig_pig_aorta}
\end{figure}

Using a Tinus Olsen tensile machine, we conducted tensile tests on several pig aortas, and found that their Gent and Fung parameters did indeed belong to those ranges, see Figure \ref{fig_pig_aorta} for  an illustrative example.

Note that the Gent and the Fung models are both connected to the so-called 
\emph{neo-Hookean model},
\be \la{neo}
W = \mu(\lambda_1^2 + \lambda_2^2 + \lambda_3^2 - 3)/2,
\en 
in the limits $J_m = \infty$, $b = 0$, respectively.

In this connection we mention the so-called \emph{third-order elasticity model}, often used to investigate the 
onset of nonlinear elasticity effects. 
When the general strain energy density of an incompressible solid is expanded in terms of powers of 
$\vec{E}$, the Green strain tensor, it is found that at the third-order of truncation (Ogden, 1974; Hamilton et al., 2004),
\be \la{third_order}
W = \mu \:\text{tr}\left(\vec{E}^2\right) + \thirdt \mathcal{A} \:\text{tr}\left(\vec{E}^3\right),
\en
where $\mu$ is the infinitesimal shear modulus of second-order (linear) elasticity, and $\mathcal{A}$ is a Landau third-order elastic coefficient. 
Turning to a representation in terms of the principal stretches, it is found that at the 
same order,
\be \la{MR}
W = \left(2\mu + \frac{\mathcal{A}}{4}\right)\left(\lambda_1^2+\lambda_2^2+\lambda_3^2-3\right)
-  \left(\mu + \frac{\mathcal{A}}{4}\right)\left(\lambda_1^2\lambda_2^2 + \lambda_2^2\lambda_3^2 + \lambda_3^2\lambda_2^2 - 3 \right).
\en
 For a plane strain deformation such as pure bending, $\lambda_3=1$ at all times
 and then $\lambda_2 = \lambda_1^{-1}$ by incompressibility. 
 In that case, comparison of \eqref{neo} and \eqref{MR} shows that  the third-order elasticity model and the neo-Hookean model coincide.
Note that in finite elasticity, \eqref{MR} is the so-called \emph{Mooney-Rivlin strain-energy density}, often used to describe rubbers.


\subsection{Pure bending of a block}
\la{pure_bending_of_a_block}


We start with a straight block of 
thickness $2 A$, width $2 L$, and height $H$,
located in the region
\be
-A \le X_1 \le A, \qquad 
-L \le X_2 \le L, \qquad 
0  \le X_3 \le H, 
\en
where $X_1$, $X_2$, $X_3$ are the coordinates in the rectangular coordinate system aligned 
with the edges of the block, giving the position of a material point in the reference configuration.
 
By applying surface tractions on the the faces at $X_2 = \pm L$, we bend the block using the deformation (Green and Zerna, 1954)
\be \la{large}
r = \sqrt{d + 2X_1/\omega}, \qquad
\theta = \omega X_2, \qquad
z = X_3,
\en
where $d$ is a constant (to be determined later), $\omega$ is a prescribed constant,  
and  $r$, $\theta$, $z$ are the cylindrical coordinates, giving the position of a material point in the deformed configuration.
The \emph{bending angle} is $\varphi \equiv 2 \omega L$.
The bent block is confined to the region 
\be
r_a \equiv  \sqrt{d - 2A/\omega} \le r \le r_b \equiv  \sqrt{d + 2A/\omega}, \quad 
- \omega L \le \theta \le \omega L, \quad 
0  \le z \le H, 
\en
where $r_a$ are $r_b$ are the radii of the inner and outer faces.

\begin{figure}
\centering \mbox{\subfigure{\epsfig{figure=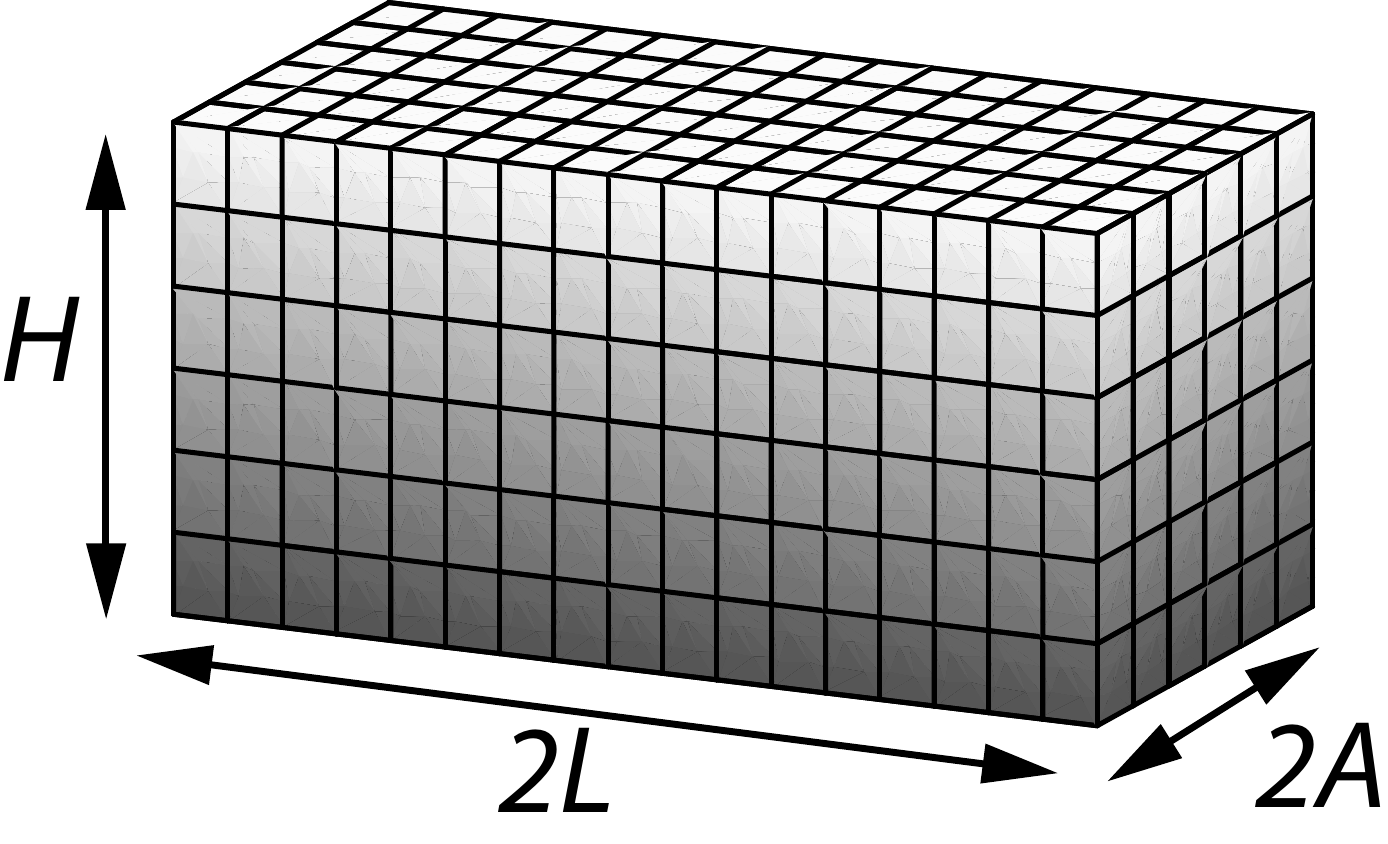, width=.48\textwidth, height = .3 \textwidth}}}
  \quad \quad
     \subfigure{\epsfig{figure=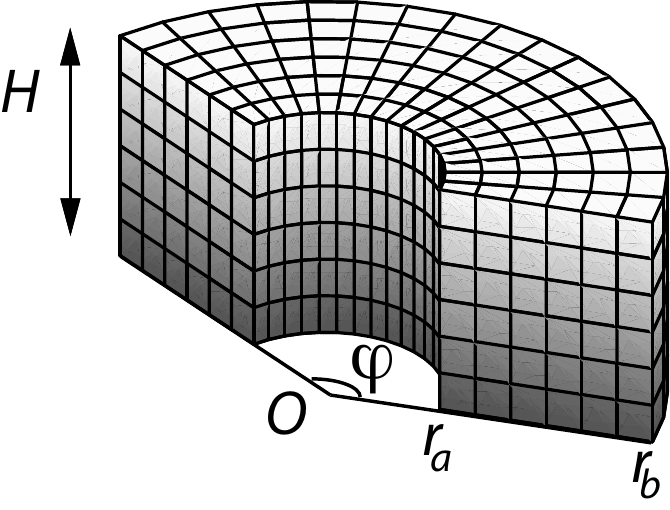, width=.45\textwidth,  height = .39 \textwidth}}
 \caption{Pure bending of a thick block. Here the original length-to-thickness ratio is 2.0 and the bending angle is $\pi/3$.
 The circumferential line elements on the outer bent face are extended during bending; those on the inner face are contracted. 
 Eventually, at a critical stretch of contraction, the inner face buckles, and axial wrinkles appear.} 
 \label{fig_bending}
\end{figure}

The corresponding principal stretches are 
\be
\lambda_1 = (\omega r)^{-1}, \qquad 
\lambda_2 = \omega r, \qquad 
\lambda_3 =1, 
\en
showing that this is a \emph{ plane strain} deformation.
These expressions also show that the strain energy density $W(\lambda_1, \lambda_2, \lambda_3)$
is a function of only one variable. 
Choosing it as $\lambda=\lambda_1$, we call $\Sigma$ the function defined by
\be
\Sigma(\lambda) = W(\lambda, \lambda^{-1}, 1).
\en
Then Rivlin (1949) shows that the principal stress component $\sigma_{1}$ is given by
\be \la{sigma11}
\sigma_{1} = \Sigma(\lambda) + K,
\en
where the constant $K$ is to be determined from the boundary conditions. 

We assume that the bent faces at $r = r_a$ and $r=r_b$ are free of normal tractions
\be \la{BC}
 \sigma_{1}(r_a) = \sigma_{1}(r_b) = 0.
 \en
These two boundary conditions lead to the determination of the constants $d$ in \eqref{large} and $K$ in \eqref{sigma11} for the strain energy densities \eqref{gent}, \eqref{fung}, and \eqref{neo}.

Take for instance the neo-Hookean form \eqref{neo}:
the boundary conditions read
\be
\mu [(\omega r_a)^{-2}  + (\omega r_a)^2 - 2]  + 2 K= 0, \qquad 
 \mu [(\omega r_b)^{-2}  + (\omega r_b)^{2} -2] + 2 K =0, 
\en
from which follows that (Green and Zerna, 1954) 
\be \la{cond_d}
\omega^2 r_a r_b = 1.
\en
Then we find $d$ and $K$ as
\be \la{d}
d = (1/\omega^2)\sqrt{1 + 4\omega^2 A^2}, \qquad 
K = -(\mu/2)[ \omega^2(r_a^2 + r_b^2) -2],
\en
and the normal stress as 
\be
\sigma_{1}(r) = \mu \omega^2 \dfrac{(r^2 - r_a^2)(r^2 -
r_b^2)}{2 r^2}.
\en
For the Gent and Fung models, $d$ is also given by \eqref{d}$_1$, whilst $\sigma_1(r)$ is equal to 
\be
-\dfrac{\mu J_m}{2} \ln\left[\dfrac{J_m+2-\frac{r_a r_b}{r^2}-\frac{r^2}{r_a r_b}}{J_m+2-\frac{r_a}{r_b}-\frac{r_b}{r_a}}\right],  \text{ and } 
\dfrac{\mu}{2b} \left[ \ee^{b\left(\frac{r^2}{r_a r_b}+\frac{r_a r_b}{r^2}-2\right)} - \ee^{b\left(\frac{r_b}{r_a}+\frac{r_a}{r_b}-2\right)}\right],
\en
respectively, 
see Kanner and Horgan (2008) and Demiray and Levinson (1982) for details.

To complete the picture, we give expressions for $\sigma_{2}$, the principal Cauchy stress normal 
to the surfaces $\theta =$ const., and for its moment $\mathcal{M}$ of the stresses on the faces 
$\theta = \pm \omega L$ about the origin:
\be
\sigma_{2} = \text{d}(r\sigma_{1} )/\text{d}r, \qquad
\mathcal{M} = H\int_{r_a}^{r_b} r \sigma_{2} \text{d}r,
\en
respectively.


\section{Bending instability}


Instability in bending is triggered by the apparition of wrinkles/creases
on the inner face of the bent block. 
Their existence is governed by the incremental equations of equilibrium and 
of incompressibility (Ogden, 1984),  
\be \la{equi}
\text{div } \vec{\dot{s}} = \mathbf{0}, \qquad 
\text{div } \vec{u} = 0,
\en
where $\vec{\dot{s}}$ is the incremental nominal stress
and $\vec{u}$ is the incremental mechanical displacement,
and by the satisfaction of appropriate boundary conditions. 

In the present context of pure bending in nonlinear elasticity, these equations 
have been written down several times (see (Coman and Destrade, 2008) and references therein), usually as a second-order system of coupled differential equations for the components of the displacement. 
Instead, we present them here as a first-order system for the displacement and the traction:
this is the so-called \emph{Stroh formulation}, which proves to be optimal for the subsequent numerical resolution of the boundary value problem. 
Omitting the details, we find that if the components $u$, $v$, $w$ of the mechanical displacement and  $\dot{s}_{rr}$, $\dot{s}_{r\theta}$, $\dot{s}_{rz}$ of the traction are in the form
\be \la{soln}
\{u, v, w, \dot{s}_{rr}, \dot{s}_{r\theta}, \dot{s}_{rz}\} = \Re\left\{\left[U(r), V(r), 0, S_{rr}(r), S_{r\theta}(r), 0\right] \ee^{\ii n\theta}\right\}, 
\en
where $U$, $V$,  $S_{rr}$, $ S_{r\theta}$, are complex functions of $r$ only, then two equations in \eqref{equi} are identically satisfied and there remains four independent equations, which can indeed be put in the Stroh form.
The constant $n$ in \eqref{soln} is the circumferential number, and is determined from the condition that there are no incremental normal tractions on the end faces $\theta = \pm \omega L$: this happens when  (Haughton, 1999)
\be
n = p \pi / (\omega L),
\en
for some integer $p$, which we call the \emph{mode number}.
Hence, the solution \eqref{soln} describes the $p$ creases appearing on the inner surface of the bent block.

Introducing the four-component displacement-traction vector (Shuvalov, 2003)
\be \la{eta}
\vec{\eta} \equiv [U, V, \ii r S_{rr}, \ii r S_{r \theta}]^t,
\en
we find that the incremental equations of equilibrium can be arranged as
\be \la{stroh}
\dfrac{\text{d}}{\text{d} r} \vec{\eta}(r) = \dfrac{\ii}{r} \vec{G}(r) \vec{\eta}(r).
\en
Here the matrix $\vec{G}$ has the following Stroh structure,
\be \label{G}
\vec{G} = \begin{bmatrix} 
  \ii & -n & 0 & 0 \\
  -n (1 - \sigma_1/\alpha ) &  - \ii (1 - \sigma_{1}/\alpha) & 0  &  -  1/\alpha\\
  \kappa_{11} & \ii \kappa_{12} & -\ii & -n (1 - \sigma_{1}/\alpha) \\
  - \ii \kappa_{12} & \kappa_{22}& -n &  \ii (1 - \sigma_{1}/\alpha)  
          \end{bmatrix}
\en
where
\begin{align}
& \kappa_{11} = 2(\beta + \alpha - \sigma_{1}) 
  + n^2\left[\gamma - (\alpha - \sigma_{1})^2/\alpha\right],
 \notag \\[2pt]
& \kappa_{12} = n \left(2 \beta + \alpha + \gamma - \sigma_{1}^2/\alpha\right),
 \notag \\[2pt]
& \kappa_{22} =  
   \gamma - (\alpha - \sigma_{1})^2/\alpha
   + 2 n^2(\beta + \alpha - \sigma_{1}),
\end{align}
and the quantities $\gamma$, $\alpha$, and $\beta$ are given in turn by (Dowaikh and Ogden, 1990)
\be \label{gammaBeta}
 \gamma =
  \lambda \Sigma'(\lambda)/(\lambda^4 - 1),
\qquad
\alpha = \lambda^4 \gamma,
\qquad
 2 \beta = \lambda^2 \Sigma''(\lambda) - 2\gamma.
\en

In the case of the neo-Hookean model \eqref{neo} and thus in the case of the third-order elasticity model \eqref{third_order} in pure bending, these quantities are 
\begin{equation}
\gamma  = \mu \lambda^{-2},
\qquad
 \alpha= \mu \lambda^{2},
 \qquad  
   2\beta = \mu(\lambda^2 + \lambda^{-2}).
\end{equation}
These quantities  have already been computed in (Destrade and Scott, 2004; Destrade and Ogden, 2005) for the Gent material in general.
Specialized to the case of plane strain they reduce to:
\begin{align}
& \alpha = \dfrac{\mu J_m \lambda^2}{J_m + 2 - \lambda^2 - \lambda^{-2}},
\qquad
\gamma = \dfrac{\mu J_m \lambda^{-2}}{J_m + 2 - \lambda^2 - \lambda^{-2}},
\notag \\[2pt]
& \beta = \dfrac{\mu J_m}{2(J_m + 2 - \lambda^2 - \lambda^{-2})}
 \left[ \lambda^2 + \lambda^{-2} + \dfrac{2(\lambda^2 - \lambda^{-2})^2}{J_m + 2 - \lambda^2 - \lambda^{-2}} \right].
 \notag
\end{align}
For the Fung material in plane strain they are:
\begin{align}
& \alpha = \mu \lambda^2 \ee^{b(\lambda^2 + \lambda^{-2}-2)},
\qquad
\gamma = \mu \lambda^{-2} \ee^{b(\lambda^2 + \lambda^{-2}-2)},
\notag \\[2pt]
& \beta = \dfrac{\mu }{2}[\lambda^2 + \lambda^{-2} + 2b(\lambda^2 - \lambda^{-2})^2]\ee^{b(\lambda^2 + \lambda^{-2} -2)}.
 \notag
\end{align}

Now if the incremental equations of motion can be solved, subject to the following boundary conditions of traction-free inner and outer faces:
\be \label{BCinc}
S_{rr} = S_{r\theta} = 0 \qquad \text{at} \qquad r=r_a, r_b,
\en
then an infinity of equilibrium states exists adjacent to the large bending, signaling the onset of instability (in the linearized sense).
When this solution is determined, we call $\lambda_\text{cr} = \omega r_a$ the value of the azimuthal stretch $\lambda_2$ on the inner face. 
It is the \emph{critical stretch of contraction} of azimuthal line elements.
Simple calculations show that 
\be
\omega A = (\lambda_\text{cr}^{-2} - \lambda_\text{cr}^2)/4,
\en
which allows for the complete determination of the current (deformed) geometry, just prior to instability.
In particular, the angle of bending, and inner and outer radii follow as
\be \label{angle_r}
\varphi = 2(\omega A)\dfrac{L}{A}, \qquad 
\dfrac{r_a}{A} = \dfrac{\lambda_\text{cr}}{\omega A}, \qquad
\dfrac{r_b}{A} = \dfrac{1}{\omega A \lambda_\text{cr}},
\en
respectively.
Note that if it turns out that $\varphi \ge 2 \pi$, then the conclusion is that the block can be completely bent into a cylinder without encountering any instability phenomenon.

                                              
\section{Numerics}


The direct numerical resolution of the differential system \eqref{stroh} is not easy to implement because of numerical stiffness issues, especially for thick blocks.

First, we use the \emph{compound matrix method} (Haughton and Orr, 1997) which is rendered optimal here by the use of the Stroh formulation.
It gives direct access to the azimuthal critical stretch of contraction.

Another option is to integrate numerically the non-linear Riccati equation satisfied by the \emph{impedance matrix}, see (Biryukov, 1985).
Here we use that integration to compute the full mechanical fields inside the bent block.


\subsection{Critical stretch}


Let $\vec{\eta}^{(1)}(r)$ and $\vec{\eta}^{(2)}(r)$ be two linearly 
independent solutions to the equations of equilibrium \eqref{stroh}.
Use them to calculate the following six \emph{compound variables} $\phi_i$,
\begin{align}
& \phi_1 =  \begin{vmatrix}
          \eta_1^{(1)} &  \eta_1^{(2)} \\
          \eta_2^{(1)} &  \eta_2^{(2)} 
         \end{vmatrix}, 
&& \phi_2 = \begin{vmatrix}
          \eta_1^{(1)} &  \eta_1^{(2)} \\
          \eta_3^{(1)} &  \eta_3^{(2)} 
         \end{vmatrix}, 
&& \phi_3 = \ii \begin{vmatrix}
          \eta_1^{(1)} &  \eta_1^{(2)} \\
          \eta_4^{(1)} &  \eta_4^{(2)} 
         \end{vmatrix}, 
\nonumber \\[4pt]
& \phi_4 = \ii \begin{vmatrix}
          \eta_2^{(1)} &  \eta_2^{(2)} \\
          \eta_3^{(1)} &  \eta_3^{(2)} 
         \end{vmatrix}, 
&& \phi_5 = \begin{vmatrix}
          \eta_2^{(1)} &  \eta_2^{(2)} \\
          \eta_4^{(1)} &  \eta_4^{(2)} 
         \end{vmatrix}, 
&& \phi_6 = \begin{vmatrix}
          \eta_3^{(1)} &  \eta_3^{(2)} \\
          \eta_4^{(1)} &  \eta_4^{(2)} 
         \end{vmatrix}.
\end{align}
Now compute their derivatives with respect to $r$ and find that they satisfy 
\be \la{CM}
\vec{\phi}' (r) = \dfrac{1}{r} A(r) \vec{\phi}(r), \qquad \text{where} \quad   \vec{\phi} \equiv
 [\phi_1, \phi_2, \phi_3, \phi_4, \phi_5, \phi_{6}]^t,
\en 
and $A(r)$ is called the \emph{compound matrix}.
Here we find that its non-zero entries are 
\begin{align}
& A_{21} = A_{51} =  A_{65}  =  A_{62} = -\kappa_{12},
&&A_{66} =  -A_{11} =  \sigma_{1}/\alpha,
\nonumber \\[4pt]
& A_{42} = -A_{23} =  A_{45} = -A_{53} = n(1 - \sigma_{1}/\alpha),
&& A_{41} = A_{63} = \kappa_{11},
\nonumber \\[4pt]
& A_{32} = -A_{24} = A_{35} =  -A_{54} = n,
&&A_{13} = A_{46} =  -1/\alpha,
\nonumber \\[4pt]
& A_{44} = -A_{33} =  2 - \sigma_{1}/ \alpha,
&& A_{31} = A_{64} = -\kappa_{22}.
\end{align}

At $r=a$, we impose $S_{rr} = S_{r\theta} = 0$ (inner face free of incremental traction), so that $\phi_2 = \phi_3 = \phi_4 = \phi_5 = \phi_6 =0$ there, and only $\phi_1$ is non-zero: this is the \emph{initial value},
\begin{equation} \la{initial}
 \vec{\phi}(a) = \phi_1(a)[1, 0, 0, 0, 0, 0]^t. 
\end{equation}
At $r=b$, the tractions must also vanish so that $\phi_6$ must be zero there: this is the \emph{target condition},
\begin{equation} \la{target}
 \phi_{6}(b) = 0.
\end{equation}

Integrating numerically the initial value problem \eqref{CM}-\eqref{initial} poses no difficulty. 
Note that the Stroh formulation yields a most simple and optimal form for the elements of the compound matrix (other formulations, for example (Haughton, 1999; Coman and Destrade, 2008) involve derivatives of the elastic moduli with respect to $r$). 

Once the dimensions of the block are given, and its strain energy density is fixed, it remains to adjust $\lambda$ so that \eqref{target} is satisfied. 
This search yields the \emph{critical stretch of contraction} for circumferential line elements on the inner face and by extension, the critical angle of bending, see previous section. 
Note that to each mode number $p$ corresponds a different value of the critical stretch;
however only the highest value is meaningful, as the others cannot be reached once the buckling has occurred. 

\begin{figure}
\center
\epsfig{figure=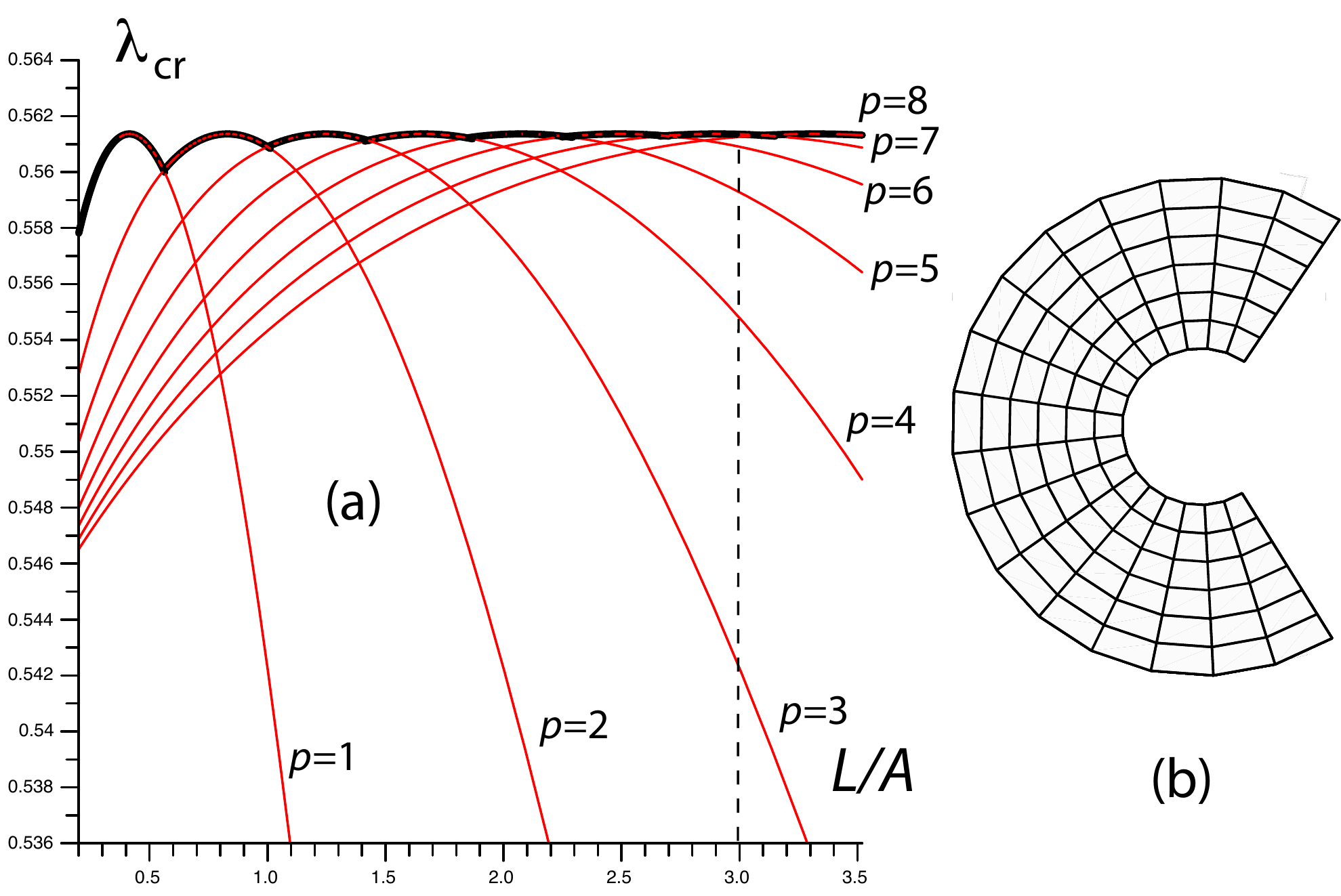, width=.9\textwidth}
 \caption{{Bending instability of rubber: for the neo-Hookean model, the Mooney-Rivlin model, and the general incompressible model of third-order elasticity, the circumferential line elements on the inner face can be contracted by 44\% at most. 
 Each different plot corresponds to a different mode, here the first eight. 
 Each plot can be deduced from the first one by a scaling factor.
 Figure (a) reveals that a block with aspect ratio $L/A = 3.0$, say, is going to buckle in mode $p=7$. Figure (b) shows that bent block just prior to buckling. }
 }
 \label{fig_neo_only}
\end{figure}

Take for instance the neo-Hookean model \eqref{neo}.
Figure \ref{fig_neo_only} shows the dependence of the critical stretch on the width-to-thickness ratio $L/A$. 
Each curve corresponds to a different mode number, but only a small part near its maximum is relevant, as highlighted by the thicker stroke. 
To generate Figure \ref{fig_neo_only}, the equations were non-dimensionalized, and in the end, only $\lambda_\text{cr}$ and $L/(pA)$ remained as unknowns. 
Then each curve in the figure is drawn by taking $p=1,2,3,\ldots$ in turn and becomes a scaled version of the $p=1$ curve. 
As a consequence, the maximum of each curve is the same. 
This scaling explains the general impression given by the figure: although ``dispersion'' is introduced due to the characteristic dimensions of the block, the value of the critical stretch itself is quite insensitive to mode numbers, and remains in the neighborhood of 0.562 approximatively (Haughton, 1999; Coman and Destrade, 2008).

For a specific example, take a block with aspect ratio $L/A=3.0$.
When it is modeled by the neo-Hookean, Mooney-Rivlin, or general third-order elasticity theory, the numerical calculations of the compound matrix method predict that it buckles in bending when $\lambda_2 = \lambda_\text{cr} = 0.5613$, in mode $p=7$, see Figure \ref{fig_neo_only}(a).
Then the formulas \eqref{angle_r} give the bending angle as $\varphi = 246^\circ$, and the inner and outer radii as $r_a = 0.78 A$, and $r_b = 2.5 A$, respectively, see Figure \ref{fig_neo_only}(b).


\subsection{Mechanical fields}


The compound matrix method is most appropriate to obtain quickly and accurately the critical stretch of contraction---the ``eigenvalue'' of the Stroh problem \eqref{stroh}.
By establishing links between the equations satisfied by the compound variables and those satisfied by the mechanical field variables, Haughton (2008) shows that it is possible to determine those latter fields throughout the bent block---the ``eigenvectors''.
Here we present a self-contained, alternative protocol for finding the ``eigenvalues'' and the ``eigenvectors'', based on the so-called \emph{impedance matrix}.
This approach can be dated back to Biryukov (1985), see Fu (2005).

First, we follow Shuvalov (2003) and call  $\vec{M}(r,r_a)$ the
\emph{matricant} solution to \eqref{stroh};
it is defined as the matrix such that 
\begin{equation} \label{M}
\vec{\eta}(r) = \vec{M}(r,r_a) \vec{\eta}(r_a), \qquad 
\vec{M}(r_a,r_a)  = \vec{I}_{(4)},
\end{equation}
and it has the following block structure
\begin{equation} \label{M_i}
 \vec{M}(r,r_a) =  \begin{bmatrix} \vec{M}_1(r,r_a) & \vec{M}_2(r,r_a)
\\[2pt] 
                    \vec{M}_3(r,r_a) & \vec{M}_4(r,r_a) \end{bmatrix},
\end{equation}
say.
We call  $\vec{S} \equiv 
[S_{r r}, S_{r \theta}]^t$ the  \emph{traction vector}, and $\vec{U}
\equiv [U, V]^t$ the \emph{displacement vector}, so that 
$\vec{\eta} = [\vec{U}, \ii r \vec{S}]^t$. 
We use the incremental boundary condition $\vec{S}(r_a) = \vec{0}$ in \eqref{M} and
\eqref{M_i} to find that
\begin{equation} \label{z}
 r\vec{S}(r) =  \vec{z^a}(r) \vec{U}(r),
 \qquad \text{where} \quad 
 \vec{z^a} \equiv - \ii \vec{M}_3 \vec{M}_1^{-1} 
   \end{equation}
is the \emph{conditional impedance matrix} (Shuvalov, 2003)
(here ``conditional'' refers to the assumed inclusion of the traction-free incremental boundary condition at $r=r_a$).

Substituting the impedance matrix $\vec{z^a}$ into the incremental equations of equilibrium \eqref{stroh} gives
\be \label{imped}
 \dfrac{\text{d}}{\text{d} r} \vec{U} =
  \dfrac{i}{r} \vec{G}_1\vec{U} -  \dfrac{1}{r} \vec{G}_2\vec{z^a U}, \qquad
\dfrac{\text{d}}{\text{d} r} (\vec{z^a U}) =
  \dfrac{1}{r} \vec{G}_3\vec{U} + \dfrac{i}{r} \vec{G}_1^+\vec{z^a U}, 
  \en
where $\vec{G}_1$, $\vec{G}_2$, $\vec{G}_3$, and $\vec{G}_1^+ \equiv \overline{\vec{G}}_1^t$ are the $2 \times 2$ sub-blocks of $\vec{G}$. 
Eliminating $\vec{U}$ between these two equations results in the following \emph{Riccati differential equation} for $\vec{z^a}$,
\begin{equation}
 \label{riccati}
 \dfrac{\text{d}}{\text{d} r} \vec{z^a} = \dfrac{1}{r} 
 \left[ \vec{z^a G}_2\vec{z^a} + \vec{G}_3 - \ii \vec{z^a G}_1 + \ii
\vec{G}_1^+ \vec{z^a} \right],
\qquad 
\vec{z^a}(r_a) = \vec{0},
\end{equation}
where the \emph{initial condition} follows from \eqref{z}$_2$ and \eqref{M}$_2$.
Note that this Riccati equation is \emph{real} because $\vec{G}_2$ and $\vec{G}_3$ are Hermitian, see \eqref{G}, and so is $\vec{z^a}$ (Shuvalov, 2003).

Now integrate \eqref{riccati} numerically, and adjust the azimuthal stretch ratio so that the incremental boundary condition of a traction-free face is satisfied on $r=r_b$. 
In other words, find the critical value $\lambda_\text{cr}$ for $\lambda_2$ so that $\text{det}\;\vec{z^a}(r_b) = 0$.
Then $\vec{S}(r_b) = \vec{z^a}(r_b) \vec{U}(r_b) = \vec{0}$ means that
\be
\dfrac{V(r_b)}{U(r_b)} = -\dfrac{z^a_{11}(r_b)}{z^a_{12}(r_b)} = - \dfrac{z^a_{21}(r_b)}{z^a_{22}(r_b)}.
\en
This ratio determines the shape of the axial creases on the outer face of the bent block.
Note however that their amplitude is not known because the stability analysis is linear (that is, linearized in the neighborhood of a large bending).

In principle we should be able to integrate simultaneously \eqref{imped}$_1$ and \eqref{riccati}$_1$ from $r_b$ to $r$ in order to determine the displacement field. 
In practice, this computation can run into numerical difficulties, by encountering singularities (Biryukov, 1985).
Instead, we use $\vec{z^b}$, the other conditional impedance matrix, found from the condition that $\vec{S}(r_b)=\vec{0}$.
We thus integrate simultaneously 
\be \label{imped_b}
 \dfrac{\text{d}}{\text{d} r} \vec{U} =
  \dfrac{i}{r} \vec{G}_1\vec{U} -  \dfrac{1}{r} \vec{G}_2\vec{z^b U}, \qquad
\dfrac{\text{d}}{\text{d} r} \vec{z^b} = \dfrac{1}{r} 
 \left[ \vec{z^b G}_2\vec{z^b} + \vec{G}_3 - \ii \vec{z^b G}_1 + \ii
\vec{G}_1^+ \vec{z^b} \right],
\en
with initial conditions 
\be
\vec{U}(r_b) = U(r_b)\left[1, -z^a_{11}(r_b)/z^a_{12}(r_b)\right]^t, \qquad
\vec{z^b}(r_b) = \vec{0}.
\en
This procedure is robust, and it gives the entire mechanical displacement field throughout the thickness of the bent block.
Equation \eqref{z}$_1$ then gives the incremental stress field.
Alternatively, the stress field can be found from the Riccati differential equations satisfied by the conditional \emph{admitance} matrices (Shuvalov, 2003).
 
\begin{figure}
\centering \mbox{\subfigure{\epsfig{figure=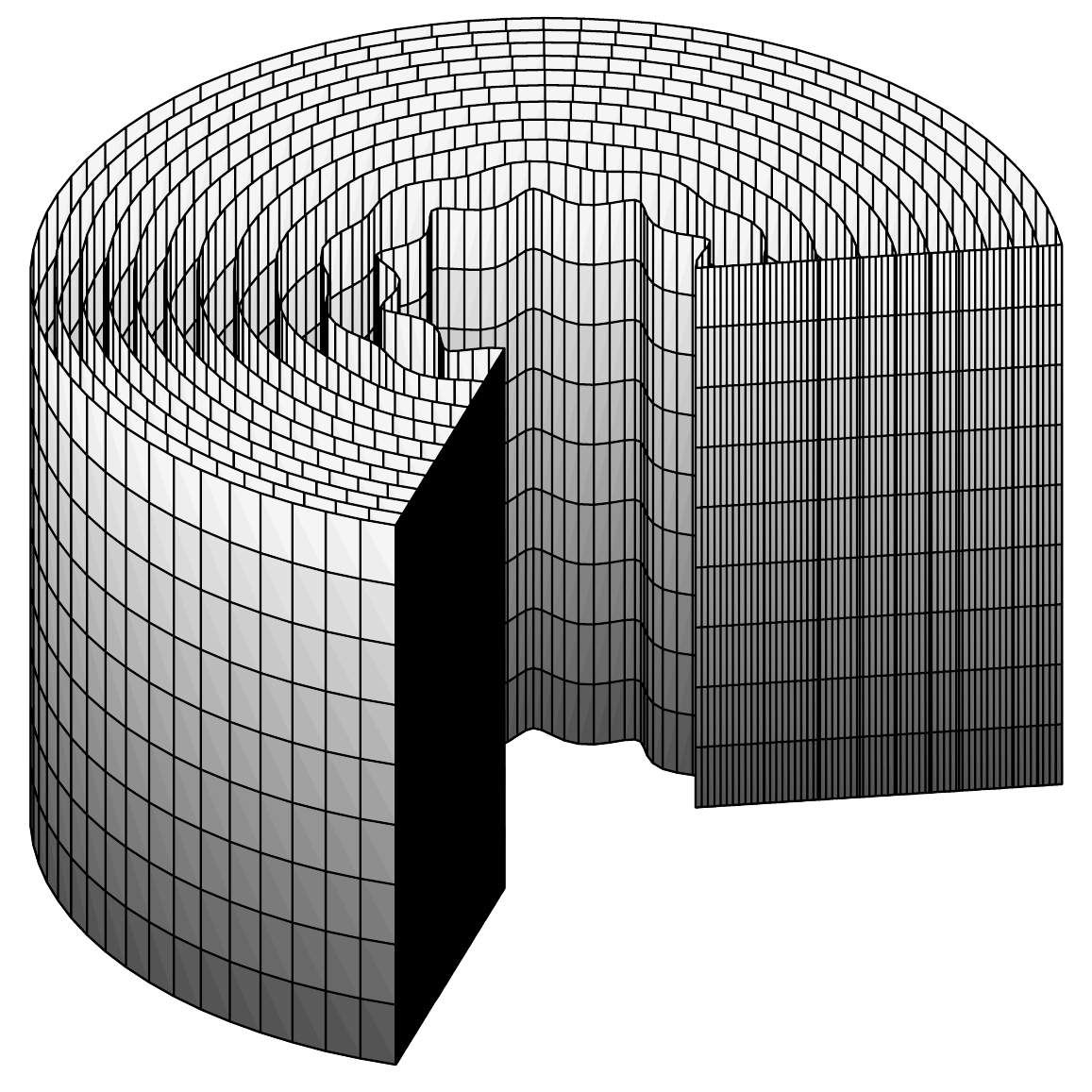, width=.5\textwidth}}}
  \quad \quad
     \subfigure{\epsfig{figure=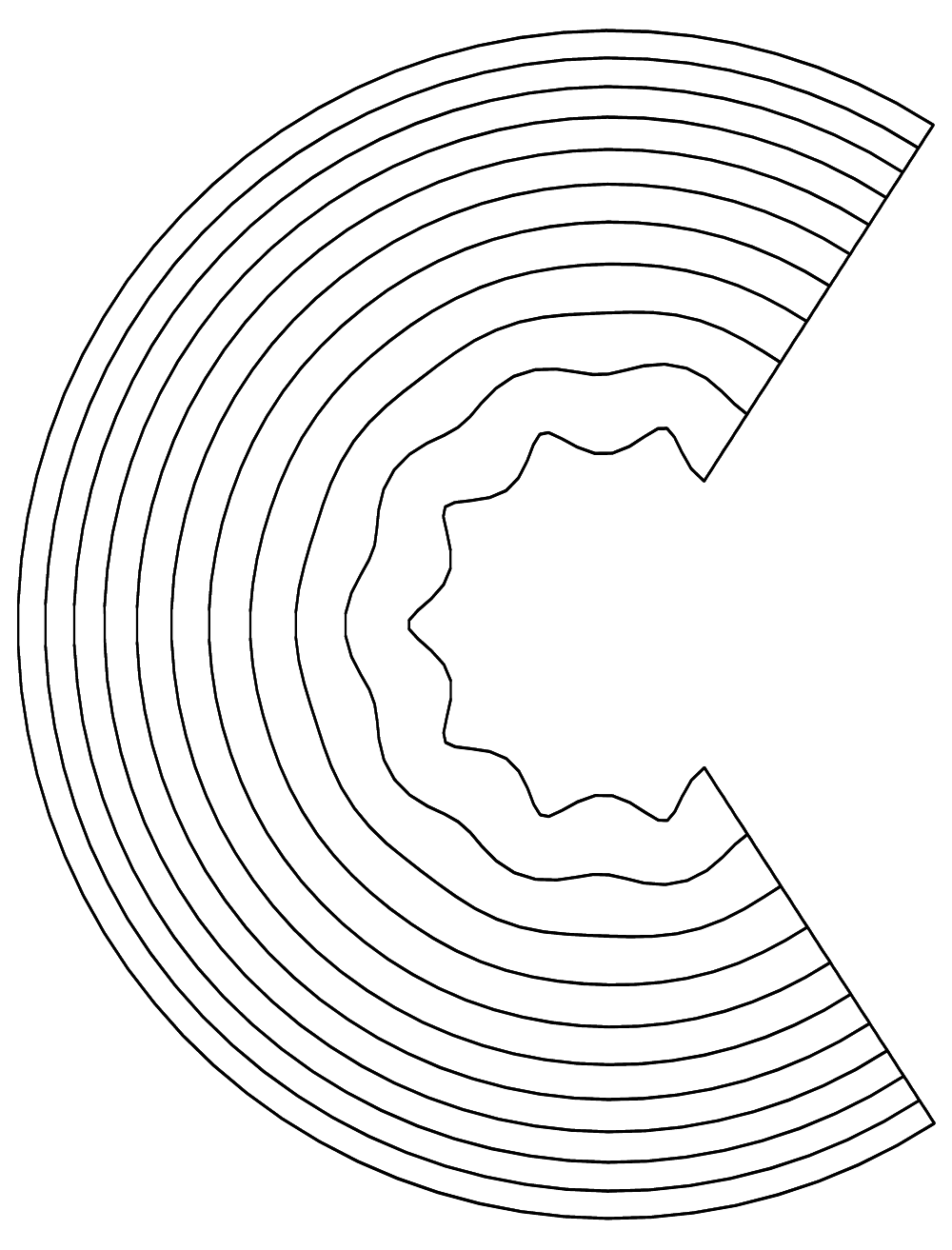, width=.4\textwidth}}
 \caption{Incremental buckling of a bent rubber block. 
 Here the solid is modeled by the neo-Hookean, Mooney-Rivlin, or general third-order elasticity solid.
 Its original length-to-thickness ratio was $L/A=3.0$. 
 The theory predicts that seven creases appear on the inner face, once the critical angle of bending is reached.}
 \label{neo_hookean_buckle}
\end{figure}

Figure \ref{neo_hookean_buckle} shows the $L/A=3.0$ rubber block of the previous section, in its incrementally buckled state. 
Ten circumferential lines are displayed, clearly showing the predicted seven axial wrinkles, and the strong localization of the displacement near the inner bent face. 
In particular, we find that the displacement amplitude on that face is more than 3000 times the amplitude on the outer face. 


\subsection{Results for strain-stiffening models}


In their study on the stability of compressed blocks, Gent and Cho (1999) conjecture that it is ``not generally necessary to consider stress-strain relations incorporating finite-extensibility effects'', at least for normal \emph{Gent rubbery materials}, for which $J_m$ lies between 20 and 200. 
Indeed, using the numerical techniques exposed in the previous sections, we find that the bifurcation curves for the Gent material \eqref{gent} at $J_m=200$ are virtually indistinguishable from those of the neo-Hookean material (which corresponds to $J_m = \infty$).
Even at $J_m = 20.0$, the curves are very close to those of the neo-Hookean solid, raising the average critical stretch ratio from 0.562 to only 0.564. 

However the situation is different for values of $J_m$ compatible with the tensile behavior of \emph{biological soft tissues}. 
Hence for $J_m=2.3$, a value measured for a young human thoracic aorta (Horgan and Saccomandi, 2003), we find that the critical stretch ratio is raised by few percents, at around 0.59. 
For $J_m = 0.4$, a value for a 70 year old human thoracic aorta (Horgan and Saccomandi,  2003), it is raised to 0.73 approximatively, showing that the earlier the finite-extensibility effects are felt, the more unstable the block is in bending.

We also find that the stiffer Gent materials buckle with less wrinkles than the softer ones. 
Take again the block with length-to-thickness ratio $L/A=3.0$.
For $J_m = \infty$, 7 wrinkles appear when $\phi = 246^\circ$, see previous section;
for $J_m=20.0$, 6 wrinkles appear, at almost the same angle of bending;
for $J_m=2.3$, we have 4 wrinkles, at $\phi = 215^\circ$;
and for $J_m=0.4$, we are down to 2 wrinkles, at $\phi = 113^\circ$.

\begin{figure}
\center
\epsfig{figure=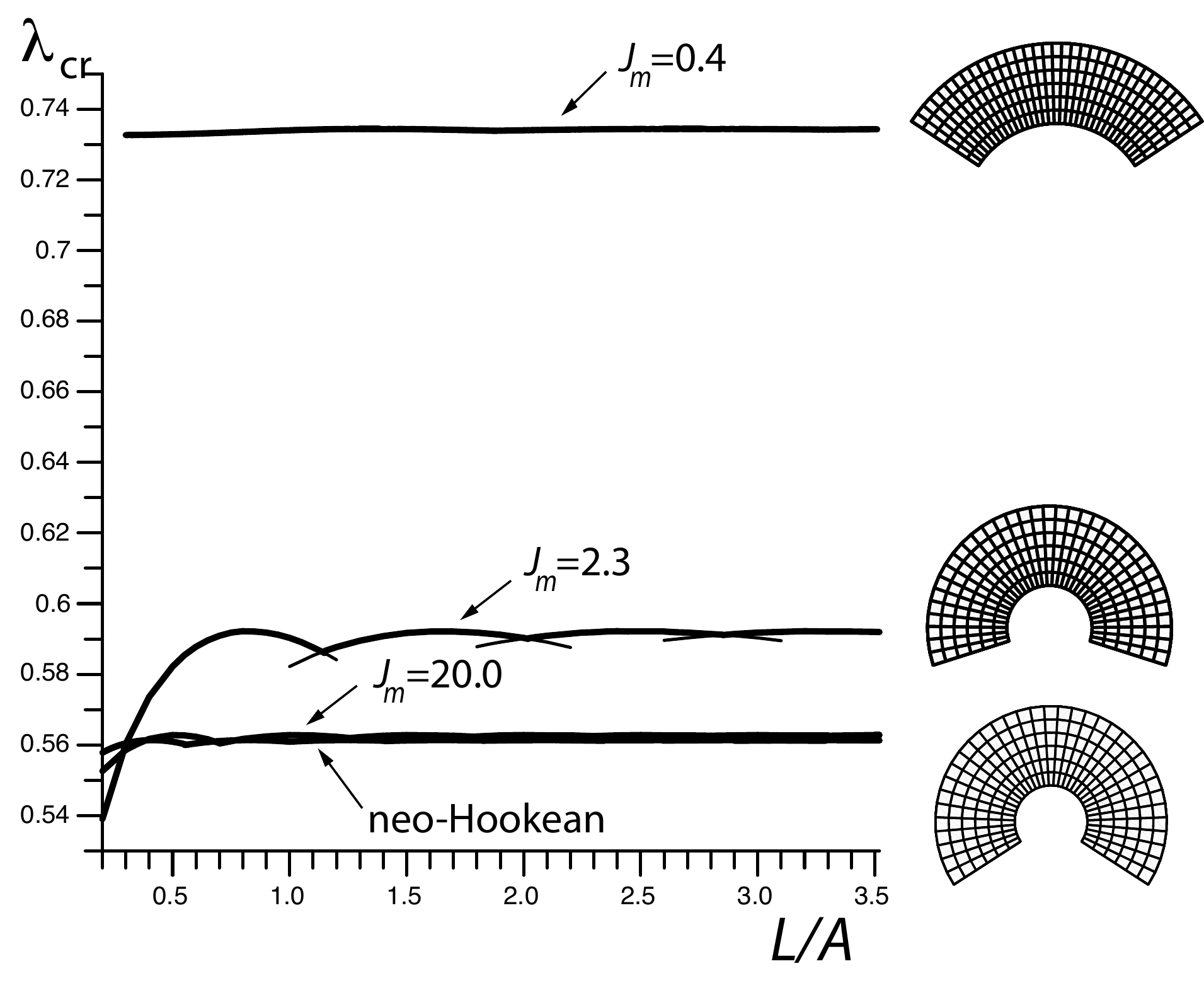, width=.9\textwidth}
 \caption{How strain stiffening affects bending instability: 
 When the solid is described by the neo-Hookean model, the Mooney-Rivlin model, the general incompressible model of third-order elasticity, or a ``rubber'' Gent model ($20 \le J_m \le 200$), the instability occurs when the circumferential line elements are contracted by 44\% ($\lambda_\text{cr} = 0.56$);
 When it is described by a ``young artery'' Gent model ($J_m = 2.3$), they can contract by 41\% ($\lambda_\text{cr} = 0.59$);
 When it is described by an ``old artery'' Gent model ($J_m = 0.4$), they can contract by 27\% ($\lambda_\text{cr} = 0.73$) only, before bending creases form.
 The figures on the right show how much a corresponding block with $L/A=3.0$ can be bent before buckling.}
 \label{fig_neo_gents}
\end{figure}

These results are reported in Figure \ref{fig_neo_gents}.

\begin{figure}
\center
\epsfig{figure=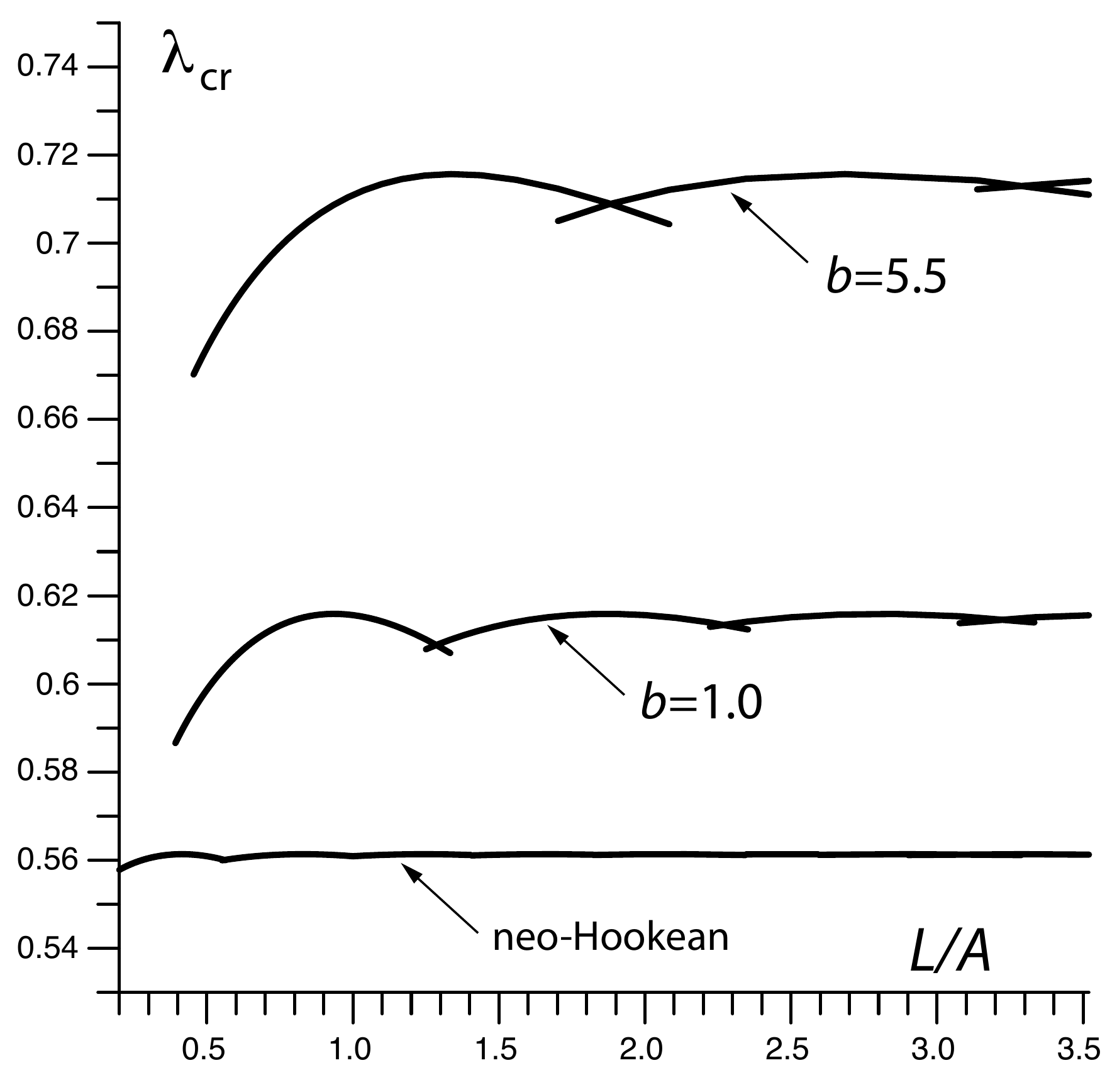, width=.9\textwidth}
 \caption{How strain-stiffening affects bending instability: 
 Here the solid is described by the Fung exponential model for arteries.
 For a ``young artery'' Fung model ($b = 1.0$), the circumferential line elements can contract by 38\% ($\lambda_\text{cr} = 0.62$);
 For an ``old artery'' Fung model ($b = 5.5$), they can contract by 28\% ($\lambda_\text{cr} = 0.72$) only, before bending creases form.
 }
 \label{fig_neo_fungs}
\end{figure}

For the \emph{Fung material} \eqref{fung}, the numerical results are remarkably similar.
Hence when the model is adjusted to account for the behavior of a 21-year-old artery, the stiffening parameter is roughly $b=1.0$, and the critical stretch ratio of contraction is about $\lambda_\text{cr} = 0.62$; 
when it is adjusted for a 70-year-old artery, we find that  $\lambda_\text{cr} = 0.72$, see Figure \ref{fig_neo_fungs}.

\begin{figure}
\centering \mbox{\subfigure{\epsfig{figure=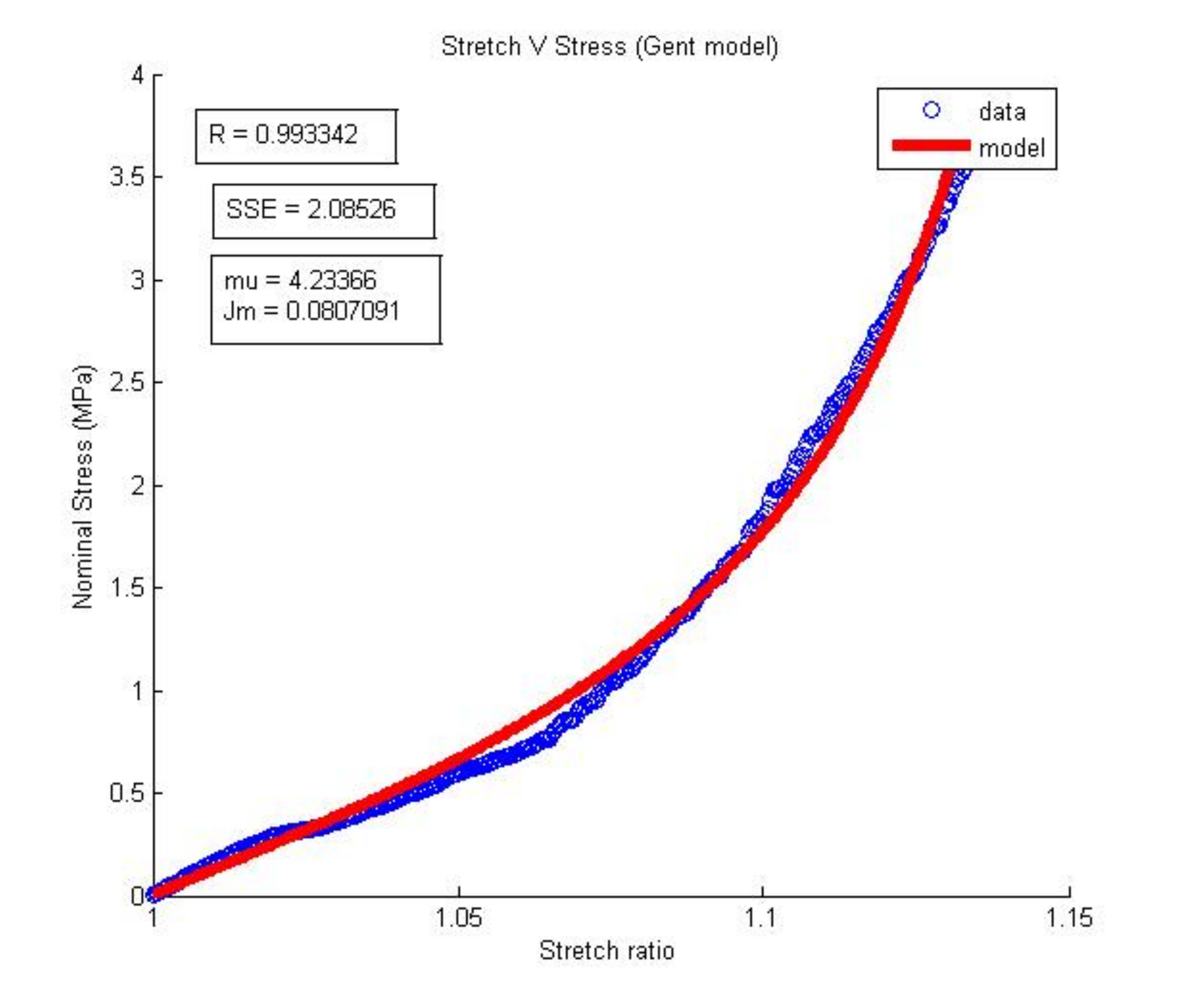, width=.43\textwidth}}}
  \quad \quad
     \subfigure{\epsfig{figure=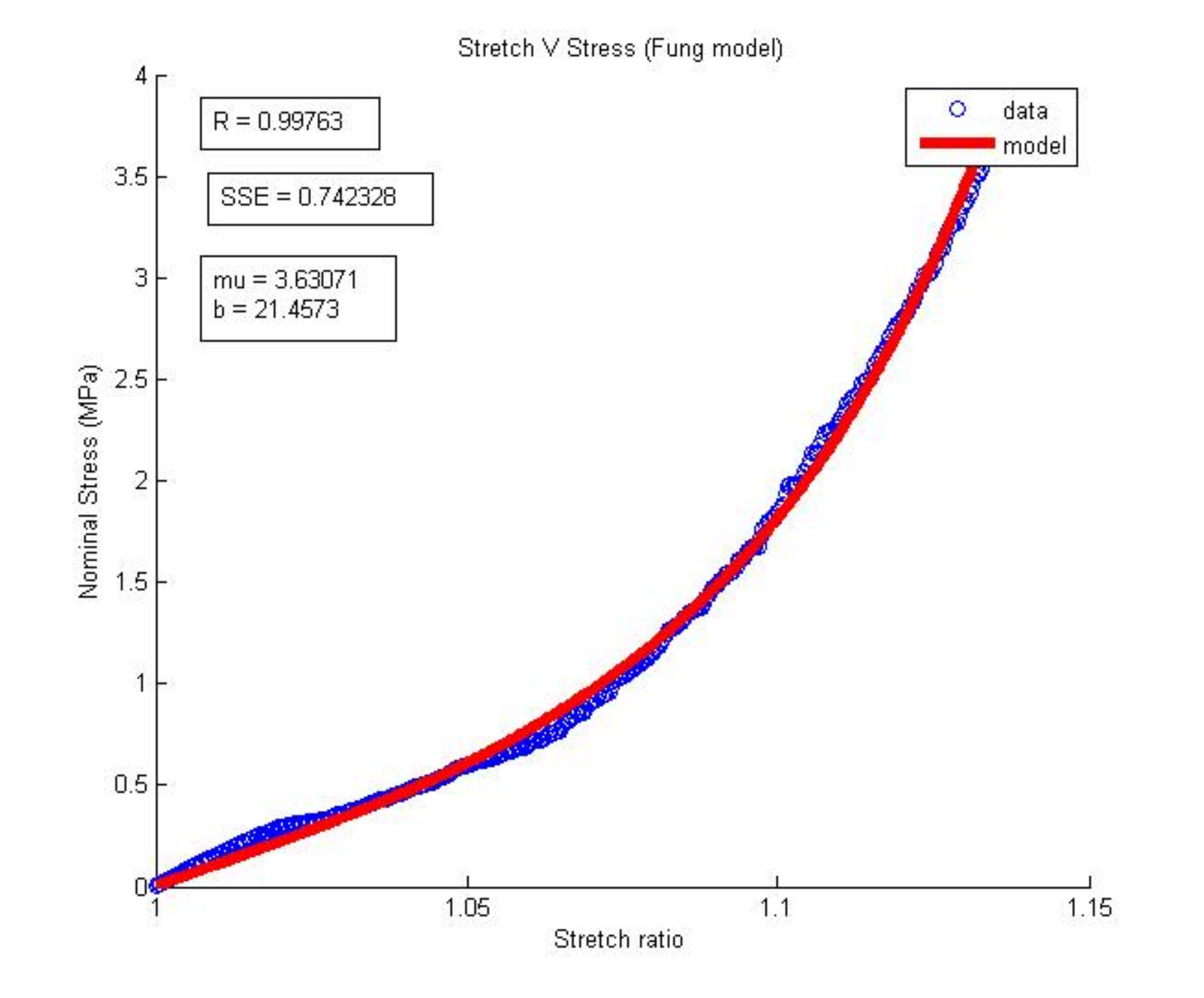, width=.45\textwidth}}
 \caption{ Tensile test on pig skin: nominal stress versus stretch ratio. 
 Experimental data and curve fitting using the Gent model (on the left) and the Fung model (on the right). 
 Curve fitting analysis gives $J_m \simeq 0.08$ with the Gent model  and $b \simeq 21.5$ with the Fung model.}
 \label{fig_pig_skin}
\end{figure}

For other, stiffer, soft tissues, the instability occurs even earlier. 
Hence we conducted tensile tests on porcine skin, and estimated that $J_m < 0.1$ when it is modeled with the Gent material, and that $b>20$ when it is modeled with the Fung material, see Figure \ref{fig_pig_skin}.
Then, the corresponding critical stretch of contraction is about $\lambda_\text{cr} = 0.80$.


\section{Concluding remarks}


The neo-Hookean material, the Mooney-Rivlin material, and consequently, the general third-order elasticity incompressible material 
are not representative of strain-stiffening solids (for instance, they do not stiffen in shear).
For these classes, often used to model rubbers or gels, the critical stretch of contraction for circumferential line elements in bending is $\lambda_\text{cr} = 0.56$. 
In contrast, the Gent and the Fung material are strain-stiffening materials \emph{par excellence}. 
When they are used to model stiff (old) arteries, the theory of incremental instability gives $\lambda_\text{cr} \simeq 0.73$. 
For pig skin, it is even raised further, to about $\lambda_\text{cr} \simeq 0.8$.
We may thus conclude that the strain-stiffening effect actually \emph{promotes} bending instability. 

Of course, these are theoretical and numerical predictions. 
Nevertheless, it must be kept in mind that the Gent and the Fung materials are popular models in the biomechanics literature and in bioengineering simulations. 
It must also be remembered that, irrespective of how advanced a Finite Element Analysis software package is, and of how precisely the image and geometry of a given soft tissue are captured, numerical simulations are, \emph{in fine}, as good ---or as bad--- as the constitutive equations they rely upon. 

If a biological soft tissue exhibits strain-stiffening effects, then these must be integrated into the constitutive model. 
If the model predicts instabilities when none are observed, then the model must be refined or abandoned. 
In fact different models react differently to different types of instabilities, and trends or classifications of models with respect to instabilities are hard to decipher in non-linear elasticity (Goriely et al., 2006).

Another major effect exhibited by biological soft tissues is that of \emph{a\-ni\-so\-tro\-py}, due to the presence of collagen fiber bundles embedded into the elastin matrix. 
The introduction of preferred directions into a constitutive model can have dramatic repercussions with respect to stability analysis (Destrade et al., 2008) and it seems crucial to be able to incorporate and quantify this effect. 
Other important differences between elastomers and soft tissues include non-linear viscoelasticity, heterogeneity, growth, and spatial distribution of fibers.


\section*{Acknowledgments}


This work was supported for the first and third author: by an International Joint Project grant
awarded by the Royal Society of London (UK) and by the  
Centre National de la Recherche Scientifique (France); for the first author:
by a Marie Curie Fellowship for Career Development awarded by the Seventh Framework Programme (European Commission); and for the second author: by a Postgraduate Research Scholarship awarded jointly by the Irish Research Council for Science, Engineering and Technology and by the Irish  Department of Justice, Equality and Law Reform.



\end{document}